

\input phyzzx

\pubtype={}
\Pubnum={IMPERIAL/TP/91-92/02}
\date={Oct. 1991}
\titlepage
\title{ ${\cal W}$-Algebra Symmetries of Generalised Drinfel'd-Sokolov
                                  Hierarchies}
\author{B. Spence}
\address{Dept. Theoretical Physics\break
         Blackett Laboratory\break
         Imperial College\break
         London SW7 2BZ.}
\abstract {Using the zero curvature formulation, it is shown that
            ${\cal W}$-algebra transformations are symmetries of
            corresponding
            generalised Drinfel'd-Sokolov hierarchies.
            This result is illustrated with the examples of
            the KdV and Boussinesque hierarchies, and the hierarchy associated
            to the Polyakov-Bershadsky ${\cal W}$-algebra. }

\endpage

\def\w{${\cal W}$}
\def\ds{Drinfel'd-Sokolov}
\def\half{{1\over 2}}
\def\third{{1\over3}}
\def\tthirds{{2\over3}}
\def\del{\partial}
\def\twomat#1#2#3#4{ \pmatrix{#1&#2\cr#3&#4\cr}}
\def\threemat#1#2#3#4#5#6#7#8#9{\pmatrix{#1&#2&#3\cr#4&#5&#6\cr#7&#8&#9\cr}}
\def\ep{\epsilon}
\def\partl#1{{\partial\over\partial{#1}}}
\def\partll#1#2{{\partial{#1}\over\partial{#2}}}
\def\reak{{\hfil\break}}

\REF\drinsok{V.G. Drinfel'd and V.V. Sokolov, J. Sov.Math. 30 (1985) 1975;
              Sov. Math. Doklady 23 (1981) No. 3, 457.}
\REF\douglas{M. Douglas, Phys. Lett. B238 (1990) 176.}
\REF\douglasa{T. Banks, M. Douglas, N. Seiberg and S. Shenker,
             Phys. Lett. B238 (1990) 279;\reak
              D. Gross and A. Migdal, Nucl. Phys. B340 (1990) 333.}
\REF\ade{E. Br\'ezin, M. Douglas, V. Kazakov and S. Shenker, Phys. Lett. B237
            (1990) 43;\reak
         D. Gross and A. Migdal, Phys. Rev. Lett. 64 (1990) 717;\reak
          \v C. Crnkovi\'c, P. Ginsparg and G. Moore, Phys. Lett. B237 (1990)
                196;\reak
           P. Di Francesco and D. Kutasov, Nucl. Phys. B342 (1990) 589.}
\REF\soton{S. Dalley, C.V. Johnson and T. R. Morris, Mod. Phys. Lett. A6
            (1991) 439, Phys. Lett. B, in press, Southampton
           preprints 90/91-16, 90/91-28;\reak
           C.V. Johnson, T.R. Morris and B. Spence, Southampton preprint
            90/91-30.}
\REF\walg{M. Fukuma, H. Kawai and R. Nakayama, Int. J. Mod. Phys. A6 (1991)
          1385, Tokyo/KEK preprints UT-572 KEK 90-165, UT-582 KEK 91-37;
          \reak R. Dijkgraaf, H. Verlinde and E. Verlinde,
           Nucl. Phys. B348 (1991) 435;\reak
          H. La, Mod. Phys. Lett. A6 (1991) 573,
              Pennsylvania preprint UPR-0453T;
          \reak H. Goeree, Nucl. Phys. B358 (1991) 737.}
\REF\hollow{M.F. de Groot, T. Hollowood and J.L. Miramontes, Princeton
            preprint IASSNS-HEP-91/19, PUPT-1251.}
\REF\classw{M.V. Saveliev, Mod. Phys. Lett. A5 (1990) 2223;\reak
            L. O'Raifeartaigh and A. Wipf, Phys. Lett. B251 (1990) 361;\reak
            L. O'Raifeartaigh, P. Ruelle and I. Tsutsui, Comm. Math. Phys. in
            press; \reak
            F. A. Bais, T. Tjin, and P. van Driel, Amsterdam preprint
             IFTA-91-04;\reak
           T. Tjin and P. van Driel, Amsterdam preprint
             IFTA-91-04.}
\REF\paulme{P. Mansfield and B. Spence, Nucl. Phys. B362 (1991) 294.}
\REF\irishnew{L. Feh\'er, L. O'Raifeartaigh, P. Ruelle, I. Tsutsui
                 and A. Wipf, Dublin preprint DIAS-STP-91-17.}
\REF\irish{J. Balog, L. Feh\'er, P. Forg\'acs, L. O'Raifeartaigh and
            A. Wipf, Ann. Phys. 203 (1990) 194.}
\REF\irishmore{J. Balog, L. Feh\'er, P. Forg\'acs, L. O'Raifeartaigh and
            A. Wipf, Phys. Lett. B227 (1989) 214, B244 (1990) 435;\reak
            J. Balog, L. Dabrowski and L. Feh\'er, Phys. Lett. B244
             (1990) 227;\reak
            L. O'Raifeartaigh, P. Ruelle and I. Tsutsui, Phys. Lett.
            B258 (1991) 359.}
\REF\gelfdik{I.M. Gel'fand and L.A. Dikii, Russian Math. Surveys
                            30:5 (1975) 77.}
\REF\gengelfdik{I.M. Gel'fand and L.A. Dikii, Funct. Anal. Appl. 10
               (1976) 259.}
\REF\polyb{A.M. Polyakov, Int. J. Mod. Phys. A5 (1990) 833;\reak
           M. Bershadsky, Princeton preprint IASSNS-HEP-90/44.}
\REF\bakasd{I. Bakas and D.A. Depireux, Maryland preprint UMD-PP91-168.}
\REF\bakasdmore{I. Bakas and D.A. Depireux, Maryland preprints
               UMD-PP91-088, UMD-PP91-111.}
\REF\das{A. Das, {\it Integrable Models}, World Scientific 1989, Chapter 9.}

\section {Introduction}

Integrable hierarchies of differential equations associated to Lie algebras
have been described by Drin'feld and Sokolov [\drinsok]. These hierarchies
have been increasingly studied in recent work in conformal field theory.
Much of this research has been spurred by the formulation of the
continuum limit of the one-matrix model in terms of the formalism associated
to the KdV hierarchy [\douglas, \douglasa]. More general matrix models,
describing two-dimensional conformal
field theories coupled to two-dimensional gravity, have also been
described
in terms of Drinfel'd-Sokolov hierarchies [\douglas-\ade].
Using this approach, a stable non-perturbative definition of
these theories has been proposed recently in ref. [\soton].

In the further study of integrable hierarchies, an important question is
that of symmetries. In this letter, it will be argued that
\w-algebra transformations are
symmetries of corresponding integrable hierarchies of differential
equations. (Note that these \w-algebra transformations are different
to the \w-algebra constraints associated to matrix models, which have
been discussed
in ref. [\walg] and follow from the $L_{-1}$ constraint, which is
related to the Galilean
invariances of hierarchies.) The argument given below that a given
\w-algebra is a symmetry algebra of the corresponding (generalised) \ds\
hierarchy can be simply summarised in words - each equation in an
integrable hierarchy of differential equations can be written
as a zero-curvature condition on a two-dimensional \lq Lax connection'.
Symmetries of the hierarchy are then the set of two-dimensional
gauge transformations which preserve the form of these Lax connections -
this symmetry turns out to be the \w-algebra transformations, with additional
conditions
fixing the dependence of the transformation parameters upon the time
variables. In the final section,
this result will be illustrated using the
hierarchies of Korteweg-deVries and Boussinesque, and the hierarchy
associated to the Polyakov-Bershadsky \w-algebra.

The \ds\ classification of integrable hierarchies has recently been
generalised and includes hierarchies corresponding
to all embeddings of $sl(2)$ in Lie algebras
[\hollow] (the \ds\ hierarchies are those associated to the
\lq principal' embeddings). There is a similar classification of
\w-algebras [\classw]. The results of this paper show how
these classifications are related - a given generalised
\w-algebra is a symmetry
algebra of the corresponding generalised \ds\ hierarchy.

\section {Hierarchy Symmetries}

Integrable hierarchies of differential equations of
the \ds\ type are neatly formulated as matrix equations [\drinsok, \hollow].
This is the \lq Lax pair' formulation, the $k$-th equation
of the hierarchy being written as
$$   \partll{L}{t_k} = \big[ P_k, L \big],                 \eqn\tassie $$
where the \lq Lax operator' $L$, and the $P_k$, are matrices
which depend on the basic hierarchy fields
$\{u^I\}$, for some index set labelled by $I$. $k$ is a non-negative integer.
The fields $u^I$ are functions of the \lq times' $t_k$; as the first
equations of the hierarchy are $(\partl{x}-\partl{t_0})u^I = 0$, the
time $t_0$ is usually identified with $x$. The pair of matrices
$(L,P_k)$ are often called the \lq Lax pair'. The Lax operator can in fact be
written as $L = \del + A$, where $\del = \partl{x}$, and $A$ is
a matrix not containing freely acting
differential operators. The Lax pair equation
\tassie\ can then be written as the condition
$$      \Big[-\partl{t_k} + P_k, \del + A\Big] = 0.        \eqn\lonnie $$
If, for each $k>0$, the coordinates $(t_k,x)$ are identified as the
coordinates of a two-dimensional space, and the pair $(-P_k,A)$ as the
components of a two-dimensional connection, then the conditions \lonnie\
are simply the requirement that the curvature of this connection vanishes.
This concise description of the equations of integrable hierarchies
will be crucial in the following. The pair $(-P_k,A)$ will henceforth be
called the \lq Lax connection', and the corresponding curvature the \lq Lax
curvature'.

The study of the symmetries of an integrable hierarchy then reduces to the
study of the transformations which preserve the zero-curvature conditions
\lonnie. Modulo global questions not addressed here, it is immediately seen
that such symmetries will be realised by those two-dimensional gauge
transformations of the Lax connection which preserve the form of the Lax pair.

Via the Miura transform, the Lax pair of \ds\ corresponds to the Lax pair
of Toda field theory [\drinsok, \hollow]. In ref. [\paulme] it was shown
that the ${\cal W}A_n$-algebra
symmetry of $A_n$ Toda field theory arises as the algebra
of gauge transformations preserving the form of the $A_n$ Toda Lax pair.
Recently [\irishnew] the relationship of this approach
to Toda theory and the \ds\ formalism has been clarified.
The Lax connection component $A$
of eqn. \tassie\ is precisely the constrained WZW
current $J$ in the \lq DS gauge' of refs. [\irishnew-\irishmore]. The WZW
field equations $\bar\del J = 0$ can be written as the zero
curvature equation $[\bar\del, \del + J] = 0$. The \w-algebra is the
algebra of
residual Ka\v c-Moody gauge transformations of the
Hamiltonian reduction of the WZW model which preserves
the constraints and gauge fixing [\irishnew,\irish]. Hence it follows
immediately that this is the algebra of gauge transformations which
preserves the form of the Lax connection component $A$ of eqns.
\tassie\ and \lonnie. For those gauge transformations to represent
symmetries of the (generalised) \ds\ hierarchy, expressed as the
zero curvature equations \lonnie, they must also preserve the form
of the other component $P_k$ of the Lax connection. This
fixes the dependence of the \w-algebra transformation parameters
upon the time variables $t_k$.

The transformations of the fields $u^I$ are those determined by the
\w-algebra, and are the same for each equation in a given hierarchy.
With the dependence of the \w-algebra transformation parameters on the
time variables $t_k$ fixed by the form-invariance of the $P_k$, it then
follows that these \w-algebra transformations are symmetries of the
entire set of equations of the generalised \ds\ hierarchy. Thus a given
(generalised) \ds\ hierarchy is invariant under the transformations
generated by the corresponding (generalised) \w-algebra.

\section {Examples}

The above argument will now be illustrated with some examples.

\subsection{Virasoro Symmetries of the KdV Hierarchy}

The $k$-th equation of the KdV hierarchy is
$$        \partll{u}{t_k}  = D_2 R_k,           \eqn\foist   $$
where $u$ is a function of $x$ and the $t_k$.
$R_k$ is the $k$-th Gel'fand-Dikii
polynomial [\gelfdik], and $D_2$ is the second Hamiltonian structure
of the KdV system
$$  D_2 = {1\over4}\del^3 - u\del - \half u'.      \eqn\secoind $$
$\del = \del/\del x$ and $u' = \del u/\del x$ in eqn. \secoind.
Using $D_2$, the Poisson bracket of two $u$'s is the
Virasoro algebra, with a central term.

In the \ds\ approach to the KdV hierarchy, the Lax operator is
$$    L =  \del + \twomat{0}{u}{1}{0} \equiv \del + A. \eqn\thoid $$
The above matrix $A$ is the $x$ component of
the Lax connection  $(-P_k,A)$. The matrix $P_k$ will be defined
in a moment. The Virasoro algebra is in fact the algebra of
two-dimensional gauge transformations which preserves the form of the Lax
connection. In the case of $A$, form invariance
means that the gauge parameter matrix
$\Lambda$ must satisfy
$$    \twomat{0}{\delta u}{0}{0} = \Lambda' + [A,\Lambda] \eqn\foith $$
for some variation $\delta u$. It can be checked that this
equation fixes $\Lambda$ to be
$$ \Lambda = -\half\twomat{-\half\ep'}{-\half\ep'' - u\ep}{\ep}{\half\ep'}
                                              \eqn\foifth $$
and the variation $\delta u$ to be
$$     \delta u = D_2\ep,                         \eqn\soixth $$
where $\ep$ is some function of $x$ and the $t_k$. The algebra of the
transformations \soixth, at fixed times $t_k$, is the Virasoro algebra
with central charge $c=1/6$.

The same manipulations may be used to find the $t_k$ connection component
$P_k$. If $P_k$ is required to satisfy the Lax pair condition
$$    \partll{L}{t_k} = [P_k,L],         \eqn\seventh $$
then one finds the solution, using the $k$-th KdV equation \foist,
$$   P_k = \half\twomat{-\half R_k'}{-\half R_k'' + uR_k}{R_k}{\half R_k'}.
                            \eqn\eighth $$
The $k$-th
KdV equation \foist\ may then be written as the zero-curvature
condition ($\del_{t_k} \equiv \del/\del t_k)$
$$     [-\del_{t_k} + P_k, \del + A] = 0.       \eqn\nineth $$

Gauge transformations preserving the form of the Lax connection component
$A$ were just seen to be the Virasoro transformations \soixth. As discussed
in the preceding section, the additional requirement that these gauge
transformations preserve the form of the other Lax connection component
$P_k$, for all $k>0$,
fixes the dependence of the gauge parameter variables - here
just $\ep$ - on the variables $t_k$.
A calculation shows that the form of $P_k$ is preserved
by the gauge transformation with parameter matrix \foifth\ if $\ep$
satisfies
$$    \partll{\ep}{t_k} = \delta R_k + \half\ep R_k'
                               - \half\ep'R_k,     \eqn\tenth $$
where $\delta R_k$ is the variation of the $k$-th Gel'fand-Dikii
polynomial $R_k[u]$ induced by the variation \soixth\ of $u$.
Note that the consistency of eqns. \tenth\ is guaranteed by the realisation
of the Virasoro transformations as gauge transformations.

Since the $k$-th KdV equation \foist\ can be written as the
condition \nineth\ for
vanishing curvature of the Lax connection, it follows that gauge
transformations
preserving this connection are symmetries of the KdV equation.
With the dependence of the parameter $\ep$ upon the variables $t_k$ determined
from eqn. \tenth, it then follows that the entire KdV hierarchy of
equations is invariant under the transformations \soixth, which
generate the Virasoro algebra. This is the Virasoro symmetry of the
KdV hierarchy. Note that, by construction, this is the full set of KdV
hierarchy symmetries which act only on the functions $u$ (the
well-known Galilean and scaling symmetries act also on the variables
$(x,t_k)$, the former additionally mixing the different equations
of the hierarchy). The Virasoro symmetry of the KdV hierarchy can
be realised as a (time dependent) canonical transformation with
generating function $\int\epsilon u$. (These comments generalise to
the other hierarchies, with the generating function being
$\int\epsilon^I u^I$.)

As examples, the first equation of the KdV hierarchy
is just $\del u/\del t_0 = -{1\over4}u'$,
which is invariant under the transformation \soixth\
if $\ep$ satisfies $\del\ep/\del t_o = -{1\over4}\ep'$.
The second equation of the KdV hierarchy is the KdV equation
itself ${\del u \over \del t_1} = -{1\over 16} u''' + {3\over8}uu'$,
and is invariant under the transformations \soixth\ if
${\del \ep\over\del t_1} = -{1\over16}\ep'''  + {3\over8}\ep'u$.
Further explicit examples may be worked out from the formula \tenth, using
eqn. \soixth\ and
the explicit forms of the polynomials $R_k$ in ref.[\gelfdik].

\subsection{\w$_3$ Symmetries of the Boussinesque Hierarchy}

The Lax operator in this case is
$$     L = \del + A = \del + \threemat{0}{0}{-u}{1}{0}{-v}{0}{1}{0},
               \eqn\another $$
with the Boussinesque hierarchy variables $u, v$. The \w$_3$ algebra
is the set of gauge transformations preserving $A$ [\irish] -
a similar
calculation to that above proves that a gauge transformation,
with parameter matrix $\Lambda$, preserves the form of $A$,
$$      \delta A = \del\Lambda + [A,\Lambda], \eqn\tryit $$
if $\Lambda$ is given by
$$ \Lambda = \pmatrix{-d'-\third g'' + \tthirds gv & -d''-\third g'''
      +\tthirds(vg)' - ug &
        \matrix{-d'''-\third g''''+\tthirds(vg)'' \cr
        -2ug'-u'g-ud\cr} \cr\noalign{\vskip14pt}
         d& -\third g'' - \third vg &
     \matrix{ -d''-\tthirds g'''-\tthirds vg'\cr
              +\third v'g - ug - vd\cr} \cr \noalign{\vskip14pt}
            g& g'+d& \tthirds g'' + d' -\third vg \cr }, \eqn\whopper $$
and if the variations of $u$ and $v$ are given by
$$  \pmatrix{\delta u\cr \delta v\cr} = D_2\pmatrix{d\cr g\cr}. \eqn\lil $$
$d$ and $g$ in eqns. \whopper\ and \lil\ are arbitrary functions of
$(t_k,x)$.
$D_2$ in eqn. \lil\ is the second Hamiltonian structure matrix,
given by
$$       D_2 = \pmatrix{\del^4 + v\del^2 + 3u\del + u'&
          \matrix{\third \del^5 - \third v\del^3 + (3u-2v')\del^2
                     \cr +(3u'-2v''-\tthirds v^2)\del +
           (u''-\tthirds v'''- \tthirds vv')\cr}   \cr\noalign{\vskip14pt}
         2\del^3 + 2v\del + v' &\del ^4 + v\del^2 + (3u-v')\del
          +(2u'-v'')\cr }.               \eqn\whoppertoo $$
Identifying $T = v$ and $W = u - \half v'$, it can be checked that this
is the matrix operator defining the usual \w$_3$ algebra, with generators
$T$ the stress-tensor and $W$ the primary spin three generator.

The $k$-th equation of the Boussinesque hierarchy can be written
$$   \partl{t_k} \pmatrix{u\cr v\cr} =  D_2\pmatrix{R_k\cr S_k\cr},
                                                  \eqn\lala $$
with $R_k$, $S_k$ generalised Gel'fand-Dikii polynomials [\gengelfdik].
If a matrix $P_k(R_k,S_k)$ is defined by
$$    P_k(R_k,S_k) = \Lambda(R_k,S_k),       \eqn\ballyhoo $$
where $\Lambda(d,g)$ is given by eqn. \whopper, then an exactly analogous
calculation to that yielding eqn. \whoppertoo\ shows that the Boussinesque
hierarchy equation \lala\ can be expressed as the vanishing
curvature equation
$$         [-\del_{t_k} + P_k, \del + A] = 0.      \eqn\flatone $$
The condition that the gauge transformation represented by the
parameter matrix $\Lambda$ preserves the form of $P$ then fixes the time
dependence of the parameters $d, g$ by the requirements
$$    \eqalign { \partll{g}{t_k} &= -\delta S_k + \Big[S_k'd-S_kg''
                 -2S_kd' - \big((R_k,S_k)\leftrightarrow(d,g)\big)\Big], \crr
     \partll{d}{t_k} &= -\delta R_k + \Big[R_k'd + S_kd'' +
       \tthirds\big(S_kg''' + S_kvg'\big)
                            - \big((R_k,S_k)\leftrightarrow(d,g)\big)
                    \Big].    \cr }           \eqn\moree $$
$\delta S_k$ and $\delta R_k$ in eqn. \moree\ are the
variations in the generalised
Gel'fand-Dikii polynomials induced by the variations \lil\ in $u$ and $v$.
Thus it has been shown explicitly
that the \w$_3$-algebra transformations \lil, with the
$t_k$ dependence of the \w$_3$-algebra parameters $g,d$ determined by
eqn. \moree, are symmetries of the Boussinesque hierarchy \lala.

\subsection{Polyakov-Bershadsky Symmetries of the Associated Hierarchy}

The simplest generalised \ds\ hierarchy which arises from a non-principal
$sl(2)$ embedding in a simple Lie algebra is the hierarchy associated
to the \w-algebra of Polyakov and Bershadsky [\polyb]. This hierarchy was
first investigated by Bakas and Depireux [\bakasd], who also studied
the zero-curvature formulation from the point of view of four-dimensional
self-dual Yang-Mills equations [\bakasdmore].
For simplicity, here only
the first non-trivial set of equations in this hierarchy will be considered.
There are four fields $U, G^+,G^-,$ and $T$, functions of $x$ and $t$,
and the equations are
$$   \eqalign{ \dot U &= G^+ - G^-,\vphantom{{1\over2}} \cr
              \dot T &= \half(G^+ + G^-)',\cr}
      \quad \eqalign { \dot G^+ &= 3U^2 - T + {3\over2}U', \cr
                       \dot G^- &= -3U^2 + T + {3\over2}U'.\cr}
             \eqn\falala $$
Following ref. [\bakasd], these can be written as the zero curvature
equation
$$    [\del_x + A_x, \del_t + A_t] = 0,          \eqn\whipit $$
with
$$   A_x = \threemat{-\half U}{0}{1}{G^+}{U}{0}{T-{3\over4}U^2}{G^-}
              {-\half U}, \quad
     A_t = \threemat{0}{1}{0}{ {3\over2}U }{0}{1}{ \half(G^+ + G^-)}
            { {3\over2}U }{0}.                      \eqn\good $$
A general gauge transformation $\delta A = \del\Lambda + [A,\Lambda]$
preserves the form of the connection $A_x$ above if $\Lambda$ takes
the form
$$  \Lambda = \threemat{a}{b}{c}{f'+cG^+{3\over2}fU}{c'-2a}{f}
               {-\half c'' + \half(bG^+ + fG^-) + c(T - {3\over4}U^2)}
               {-b'+{3\over2}bU + cG^-}{a-c'},    \eqn\andbig$$
and if the variations are given by
$$ \eqalign { \delta U &= c''-2a' + bG^+ - fG^-, \crr
              \delta G^+&= f'' + 3aG^+ + c{G^+}' + 3f'U + {3\over2}fU'
                          +{3\over2}cG^+U + f(3U^2 - T), \crr
             \delta G^- &= -b'' + 3(c'-a)G^- + c{G^-}' + 3b'U
                     + {3\over2}bU' - {3\over2}cG^-U + b(T-3U^2),\crr
                  \delta T &= -\half c''' + 2c'T + cT' - {3\over2}c'U^2
               -{3\over2}cUU' +
                   \Big({3\over2}c''-3a'\Big)U + \half b{G^+}' \crr
                    &\qquad\qquad\qquad\qquad
           + \half f{G^-}'+ {3\over2}\Big(b'G^+ + f'G^-\Big), \cr }
                                                     \eqn\andbigger $$
with $a,b,c$ and $f$ arbitrary functions of $(t_k,x)$.
The variations \andbigger\ realise
the \w-algebra of Polyakov and Bershadsky. (The functions
$b,f$ and $c$ parameterise the variations induced by the
generators $G^+, G^-$ and $T$ respectively. To obtain the canonical
form of the action of the generator $U$ as given in ref. [\polyb] one
needs to shift its parameter $a$ by a term $-{3\over4}cU$.)

Finally, the connection coefficient $A_t$ in eqn. \good\ is invariant
under the gauge transformations represented by eqn. \andbig\ if the
parameters $a,b,c$ and $f$ have time dependence determined by
$$ \eqalign { \dot a &= {3\over2} \big(b-f\big)U +
                        \half c\big(G^- - G^+\big) - f',\quad
                  \dot b  = 3a - c' + {3\over2} cU, \crr
                    \dot c & = b - f, \qquad
                        \dot f  = 2c' - 3a - {3\over2} cU.\cr}
                 \eqn\melbourne $$
The transformations \andbigger\ are thus symmetries of the equations
\falala\ of this hierarchy if the transformation parameters
satisfy \melbourne. As before, this is a consequence of the fact that these
transformations can be realised as gauge transformations, and that
the original equations can be expressed as a zero-curvature condition.
Again, from the general
argument given earlier, the transformations \andbigger\ are symmetries
of the entire hierarchy of differential equations, with the parameters
$a,b,c,f$ having $t\equiv t_1$ dependence determined by eqn. \melbourne, and
$t_k$, $(k>1)$ dependence determined by generalisations of this equation.
Details can be worked out straightforwardly from the formulation of this
hierarchy in ref. [\bakasd].

\section{Further Remarks}

It has been shown above that a given generalised
\w-algebra generates symmetries
of the corresponding generalised \ds\ hierarchy of differential equations,
and this was shown in detail in some examples.
This connection between \w-algebras and integrable hierarchies
is another intriguing interconnection in the whole area of integrable
models. Elucidation of a common underlying structure is an obvious task.
The bi-Hamiltonian structure fundamental to integrable hierarchies
has a geometric meaning (see, for example, ref. [\das]), which
may be an important
clue to the direction of further progress.

{\bf Acknowledgements}: This work was begun in the Physics Department
at Southampton University, and I would
like to thank Prof. K.J. Barnes for arranging
my tenure there, and Clifford Johnson for helpful conversations.
I would also like to thank Prof. L. O'Raifeartaigh for his hospitality
during a visit to the Dublin Institute for Advanced Studies.

\refout

\bye